\documentclass{ws-procs9x6-cpt19}
\begin{document}

\newcommand{\refeq}[1]{(\ref{#1})}
\def\etal {{\it et al.}}
%any other macros go here 

\title{GINGER}

%\address{$^1$University Department, University Name,\\
%City, State ZIP/Zone, Country}
%\address{$^2$Group, Laboratory,\\
%City, State ZIP/Zone, Country}
%\author{ Angela D. V. Di Virgilio}
%\author{Filippo Bosi, Angela D. V. Di Virgilio, Umberto Giacomelli, Andrea Simonelli, Giuseppe Terreni}
%\address{INFN Sez.~di Pisa, Largo B. Pontecorvo 3,\\
%Pisa, 56124, Italy}
\author{Angela D.V. Di Virgilio$^1$}
\address{$^1$INFN Sez.~di Pisa, Largo B. Pontecorvo 3, Pisa, Italy,}

\author{on behalf of the GINGER collaboration \footnote{Altucci, C.$^2$; Basti, A.$^3$; Beverini, N.; Bosi, F.$^1$; Carelli, G.$^3$; Ciampini, D.$^3$;
 Fuso, F.$^3$; Giacomelli, U.$^1$; Maccioni, E.$^{1,3}$; Ortolan, A.$^4$;Porzio, A.$^5$;
Simonelli, A.$^1$; Stefani, F.$^3$; Terreni,G.$^1$; Velotta R.$^1$\\
$^1$INFN Sez.~di Pisa, Largo B. Pontecorvo 3, Pisa, Italy,\\
$^2$Dept.~of Physics, Univ.\ ``Federico II" and INFN, Napoli,  via Cintia, 80126 Napoli, Italy,\\
$^3$Dept.~of Physics of the Univ.~of Pisa, Largo B. Pontecorvo 3, Pisa, italy,\\
$^4$INFN-National Lab. of Legnaro, viale dell'Universit\`a 2, I-35020, Legnaro (PD), Italy,\\
$^5$CNR - SPIN and INFN, Napoli, via Cintia, 80126 Napoli, Italy}}
\begin{abstract}
GINGER (Gyroscopes IN GEneral Relativity), based on an array of large dimension ring laser gyroscopes, is aiming at measuring in a ground laboratory the gravito-electric and gravito-magnetic effects (also known as De Sitter and  Lense-Thirrings effect), foreseen by General Relativity. The sensitivity depends on the size of the RLG cavities and the cavity losses, considering the present sensitivity, and assuming the total losses of 6 ppm, with 40m perimeter and 1 day of integration time sensitivity of the order of frad/s is attainable. The construction of GINGER is at present under discussion.
\end{abstract}
\bodymatter
\section*\\
The Sagnac effect has been discovered by Georges Sagnac more than 100 years ago, and states that: the difference of time of flight of two light beams counter-propagating inside a closed path is proportional to the angular rotation rate of the frame. In 2014 the Institute of France has organised a symposium to celebrate the 100 years, the symposium book provides a wide picture of this effect and its scientific applications.\cite{CR}    In general the Sagnac signal is generated by 
the time of flight difference between the two counter propagating waves in the closed path and is related to any non-reciprocity between the two propagation directions. The most general formula is:
\begin{equation}
    \Delta\phi = \frac{8 \pi A\Omega}{\lambda c}\cos{\theta}
    \label{fig:General0}
\end{equation}
where $\Delta\phi$ is the difference in phase of the two output waves,
A is the enclosed area, 
$\lambda$ the wavelength, $\theta$ the angle between the area versor of the ring and the rotational axis of  the angular velocity $\vec{\Omega}$, and c the velocity of light, in this case the scale factor $S_0$ is $S_0 = \frac{8 \pi A}{\lambda c}$.  So far several different probes have been utilised: light, atoms and helium super-fluid.\cite{CR}
Here we limit the discussion to the light as probe which for high sensitivity application has the obvious advantage of being insensitive to gravity variations. The closed path can be an optical fiber coil or  a ring Fabry-Perot cavity \footnote{This two concept are widely used to develop inertial navigation gyroscopes.}. 
 The devices based on resonant Fabry Perot ring cavities have Very interesting properties. They can be passive or active. Passive is when the resonant cavity is interrogated injecting light from the outside (passive ring cavity, PRC);  active when it contains an active medium 
and the device is a laser emitting two counter propagating modes, in this case it is called Ring Laser Gyro (RLG or active ring cavity ARC).\cite{RSIUlli} The output of RLG is the beat frequency $f_s$ of the two modes, which is related to the $\vec{\Omega}$
\begin{eqnarray}
f_s =S \Omega \cos{\theta}, ~S = 4\frac{A}{\lambda L} 
\label{uno}
\end{eqnarray}
where L is the perimeter and S the geometrical scale factor. This is advantageous since the frequency measurement is extremely accurate and provides large dynamic range, moreover the scale factor S can be more efficiently stabilised, since it is proportional to $A/L$; this ratio can change with time, building the apparatus in order to have this ratio close to a saddle point, the requirement on the long time stability of the geometry can be relaxed\footnote{Very low frequency signals are at the base of this kind of research, as a consequence long time stability of the apparatus is necessary.}. \\
At present, Earth based RLGs with perimeters of several meters are  by orders of magnitude the most sensitive, both for short time and long time measurements; PRC are presently approaching the nrad/s in 1 second integration time, atom based gyros have recently reached $0.3$nrad/s in 3 hours integration time, while the best RLG,\cite{RSIUlli} have sensitivity of $10$prad/s in 1 second, and can integrate for one day reaching $0.1$prad/s.\cite{RSIUlli} 
They have applications for geodesy, geophysics and for General Relativity (GR) tests,\cite{CR} as GINGERINO, a 14m perimeter RLG based on a simple mechanical structure which can be easily built and oriented at will, which runs unattended and uncontrolled for months, with a duty cycle higher than $95\%$ and sensitivity below $0.1$nrad/s in 1 s measurement. GINGER is aiming at the Lense Thirring test at the level of $1\%$,\cite{prd2011,angelo2017,angela2017} \textit{the first measurement} of a $GR$ dynamic effect of the gravitational field on 
the Earth surface (not considering the gravitational redshift).
Though not in free fall condition, it would be a direct local measurement, independent from the 
global distribution of the gravitational field and 
and not an average value,  as in the case of space experiments.  The preparatory phase has provided solutions to the scale factor stabilisation\cite{Santagata2015,Belfi2014,GP2new} and the data analysis in order to eliminate the non linearity induced by the laser dynamics\cite{angela2019}. GINGER has to push the relative sensitivity of the Earth rotation rate measurement from $3$ parts in $10^{9}$ up to $1$ part in $10^{12}$, this is equivalent to reach the sensitivity level of $10^{-16}$rad/s. 
 \footnote{It is important to note that GINGER requires a very high accuracy, and it is advantageous that its apparatus can provide two different measurement technique: it could act as active or passive, providing an important potential check of the systematics. Moreover other different laser wavelength could be used.}
 RLGs have large interest in fundamental physics to study the property of the gravito-magnetic field, and recently its utilisation for Lorenz violation has been investigated  by Jay Tasson and Max Trostel.\cite{max} 
The RLG array measures a total angular rotation rate with is the sum of the cinematic Earth rotation rate $\Omega_\oplus$, and the GR terms. Since the international system IERS measures $\Omega_\oplus$, and the deSitter term is very well known, it is possible to evaluate the Lense Thirring one. The problem of the Lense Thirring test is 
a very general one, similar to any effect induced by geophysical phenomena.  The same apparatus is suitable to observe the signal at the sidereal day frequency, which brings information for the Lorenz violation.\cite{max}

\begin{figure}
     \centering
     \includegraphics[scale=0.3]{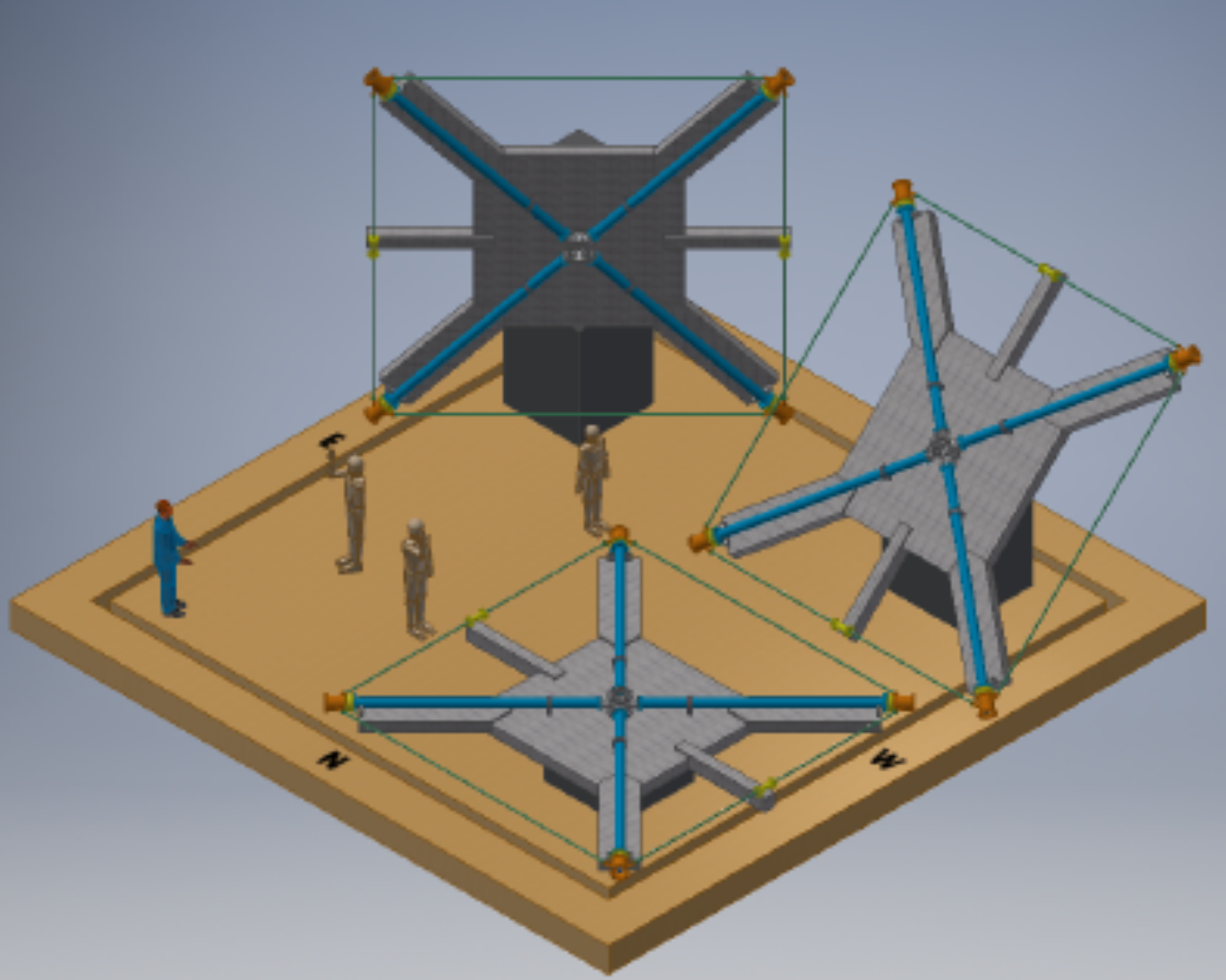}
      \includegraphics[scale=0.2,angle=90]{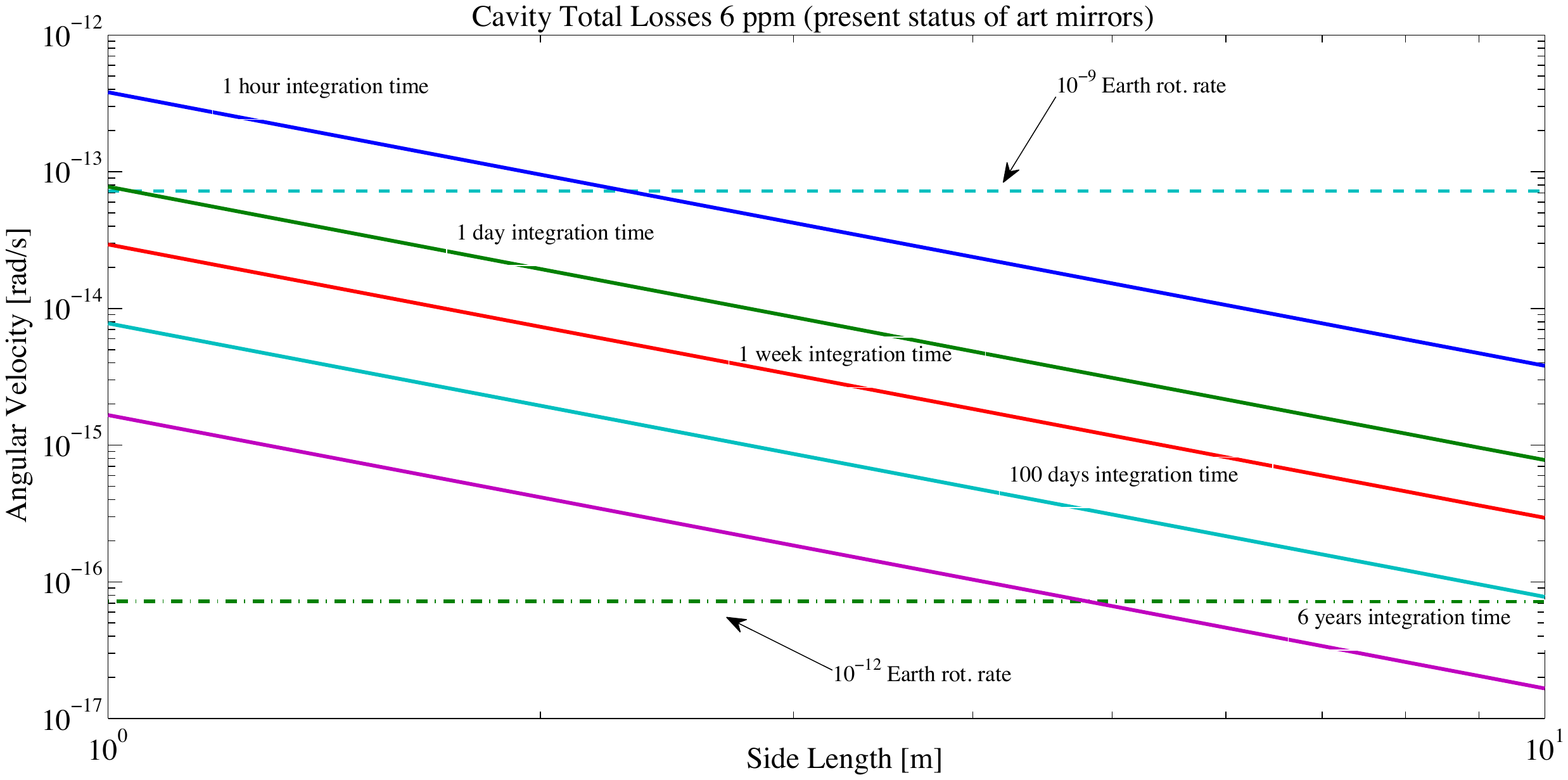}
     \caption{Pictorial view of the GINGER project, and expected sensitivity in function of side length of the square RLG of the array.}
     \label{fig:GINGER}
 \end{figure}

Fig.~\ref{fig:GINGER} shows a pictorial view  of GINGER with $3$ independent RLG and the expected sensitivity in function of the side length of square RLGs; based on the present sensitivity, it is assumed that the losses depend only on the 4 mirrors, the minimum total losses are 6ppm.\footnote{At present the status of art of mirrors is about 1.5 ppm total losses for mirror, 1 ppm scattering and 0.5 ppm transmission.} 
From all above said it is clear that RLG with 10m side could completely satisfy the GINGER target. The installation inside the Gran Sasso laboratory has been studied, where rings with side between 5 and 6m are feasible. At present we are discussing the installation  inside the  SAR-GRAV underground laboratory, which being under construction could  accept larger rings.
\bigskip\\
GINGER is a project to build an array of RLGs to detect on ground the Lense Thirring effect at $1\%$ precision; it is an INFN project, which involves the INFN sections and university of Pisa, Legnaro and Naples. Large experimental work has been pursued toward GINGER, focusing on:\cite{Santagata2015,Belfi2014,GP2new,angela2019} geometrical scale factor control, study of the optimal orientation of each RLG, and  signal reconstruction following the laser dynamic. GINGERINO, realized for test purpose inside the Gran Sasso undergorund laboratory, has shown the advantage of an underground location,\cite{90day} shielded from atmospheric and thermal variations. Presently its construction is under discussion, the underground laboratories of Gran Sasso and the new born SAR-GRAV are possible locations.

\end{document}